\documentclass[superscriptaddress,reprint,citeautoscript,prb]{revtex4-1}
\usepackage{graphicx} 
\usepackage[margin=0.7in]{geometry} 
\usepackage{amsmath}
\usepackage{hyperref} 
\usepackage{multirow}

\usepackage{newtxtext}  
\usepackage{booktabs}
\usepackage{xcolor} 
\usepackage{hyperref}
\hypersetup{colorlinks, 
	linkcolor={blue!75!black!80!yellow},
	citecolor={blue!75!black!80!yellow}, 
	urlcolor={blue!75!black!80!yellow}
	}

\makeatletter
 \renewcommand\@make@capt@title[2]{\@ifx@empty\float@link{\@firstofone}{\expandafter\href\expandafter{\float@link}}\sffamily{\textbf{#1}}\@caption@fignum@sep#2 }
\renewcommand{\Im}{\operatorname{Im}}

\renewcommand{\vec}[1]{\mathbf{#1}}
\newcommand{\sub}[1]{\ensuremath{_{\textrm{#1}}}} \newcommand{\super}[1]{\ensuremath{^{\textrm{#1}}}} \newcommand{\mobUnit}{cm\super{2}V$^{-1}$s$^{-1}$} 
\newcommand{\HarvardSEAS}{John A. Paulson School of Engineering and Applied
Sciences, Harvard University, Cambridge, MA, USA}
\newcommand{\HarvardCCB}{Department of Chemistry and Chemical Biology, Harvard
University, Cambridge, MA, USA} \newcommand{\MITPhy}{Department of Physics,
Massachusetts Institute of Technology, Cambridge, MA, USA}
\newcommand{\RPI}{Department of Materials Science and Engineering, Rensselaer
Polytechnic Institute, Troy, NY, USA}

\begin{document} 

\author{Christopher J. Ciccarino}\affiliation{\HarvardSEAS}\affiliation{\HarvardCCB}
\author{Thomas Christensen}\affiliation{\MITPhy}
\author{Ravishankar Sundararaman}\affiliation{\RPI}
\author{Prineha Narang}\email{prineha@seas.harvard.edu}\affiliation{\HarvardSEAS}

\title{Dynamics and Spin-Valley Locking Effects in Monolayer Transition Metal Dichalcogenides}

\date{\today}

\begin{abstract} 
Transition metal dichalcogenides have been the primary
materials of interest in the field of valleytronics for their
potential in information storage, yet the limiting factor has been achieving
long valley decoherence times.  We explore the dynamics of four monolayer TMDCs
(MoS\sub{2}, MoSe\sub{2}, WS\sub{2}, WSe\sub{2}) using \emph{ab initio}
calculations to describe electron-electron and electron-phonon interactions.  By
comparing calculations which both omit and include relativistic effects, we
isolate the impact of spin-resolved spin-orbit coupling on transport properties.
In our work, we find that spin-orbit coupling increases carrier lifetimes at the valence band
edge by an order of magnitude due to spin-valley locking, with a proportional
increase in the hole mobility at room temperature.  At temperatures of 50~K, we
find intervalley scattering times on the order of 100 ps, with a maximum value
$\sim$140 ps  in WSe\sub{2}. Finally, we calculate excited-carrier generation profiles
which indicate that direct transitions dominate across optical energies, even for WSe\sub{2} which has an
indirect band gap.  Our results highlight the intriguing interplay between spin
and valley degrees of freedom critical for valleytronic applications. Further, our work
points towards interesting quantum properties on-demand in transition metal dichalcogenides
that could be leveraged via driving spin, valley and phonon degrees of freedom.\\

\textbf{Keywords:} valleytronics, carrier dynamics, transition metal dichalcogenides, spin-valley locking
\end{abstract}

\maketitle

Transition metal dichalcogenides (TMDCs) represent a class of semiconducting 2D
materials of significant scientific potential.\cite{Kato:2004ve}  Specifically, TMDCs are a key
player in the fields of spintronics~\cite{Wolf:2001dq, Dankert:2017,
Han:2016} and valleytronics~\cite{Ye:2016, Xiao:2012} which seek to use degrees
of freedom beyond charge
to accelerate electronic computing and information processing. 
These materials offer quantum properties on-demand\cite{Basov:2017kx, Chumak:2015qf} 
with interesting possibilities to create topological states and non-equilibrium matter 
through driven phonon states.\cite{Hubener:2018zr, Shin:2018ly}
Monolayer TMDCs are furnished with inequivalent valleys at the $K$ and $K'$ points
of the Brillouin zone~\cite{Schaibley:2016}, due to the absence an in-plane inversion
center.  As a result, carriers in the $K$ and $K'$ valleys acquire an additional
quantum number known as their valley index or valley pseudospin. Heavy transition metal
atoms in these materials introduce strong
spin-orbit coupling, with large spin-splitting of opposite signs at the $K$ and
$K'$ valleys near the band edges, leading to spin-valley
coupling~\cite{Schaibley:2016,Yan:2017,Xu:2014, De-Giovannini:2016ys}.  Consequently, scattering of
charge carriers between valleys necessitates a simultaneous spin flip in
addition to a large momentum transfer ($K \leftrightarrow K'$), and is therefore
expected to be a slow process~\cite{Xie:2016, PhysRevLett.111.026601}. 

In the current state-of-the-art in valley physics, specific valleys can be targeted and
selectively populated via polarized light~\cite{Mak:2012, Zeng:2012} and
magnetic fields~\cite{MacNeill:2015,Aivazian:2015,Srivastava:2015}.  These
methods for breaking valley degeneracy have now been well-explored, and a
central limit in valleytronics remains the valley polarization lifetime which
directly determines the retention time of information represented by the valley
state~\cite{Kim:2017}. 

Quantifying valley polarization times in monolayer TMDCs has been led by
experimental investigation~\cite{Yan:2017,Xie:2016,Kumar:2014,Wang:2013,Mai:2014,Zhu:2014},
while theoretical work has come along only recently~\cite{Molina-Sanchez:2017}.
Difficulties in quantifying valley polarization times are clear from the wide
range (from picoseconds to nanoseconds) of reported lifetimes. 
Outside of general experimental difficulties in 2D materials, 
one reason for such a discrepancy is the role of multi-particle
excitations including excitons and trions. 
Excitons have large binding energies due to ineffective dielectric screening in 2D materials,
which complicates valley population mechanisms based on optical excitation. 
The difference between exciton valley dynamics and free 
electron or hole valley dynamics
is significant. While exciton valley lifetimes are expected to be short,
individual electron and hole lifetimes are expected to be considerably
longer, and therefore represent the best candidates for
effective information storage~\cite{Xu:2014}.

The potential for valleytronic applications is particularly enhanced
by strong spin-orbit coupling at the band edges of heavy-metal monolayer
TMDCs including MoS\sub{2}, MoSe\sub{2}, WS\sub{2} and WSe\sub{2}.
The valence band splitting is primarily due to $d_{xy}$ and $d_{x^2 - y^2}$
orbitals of the transition metal, with splits ranging up to 0.5~eV for the
heavier tungsten monolayers~\cite{Zhu:2011,Mak:2010, Zhang:2013, Zhao:2013}.
The conduction band split is significantly smaller,
as this band is mostly composed of $d$ orbitals with magnetic
quantum number $m = 0$. This small conduction band split means that the
inequivalent $K$ and $K'$ valleys are susceptible to intervalley
scattering~\cite{Molina-Sanchez:2017}. The larger splitting of the valence band
therefore makes this edge much more attractive for valleytronic applications. 
The exact role of spin-orbit coupling in carrier lifetimes and mobilities
is, however, not yet known unambiguously.

In this manuscript, we investigate valley physics and transport properties of TMDCs
using an \emph{ab initio} framework, fully including the impact of electron-electron
and electron-phonon interactions, and self-consistent spin-orbit coupling.
We compare these results to spin degenerate, non-relativistic calculations,
enabling us to quantify the crucial impact of spin-orbit coupling in increasing
the valence band lifetimes near the $K$ and $K'$ points and the corresponding hole mobilities.
In particular, we show that the large spin-orbit coupling precludes intervalley
scattering near the valence band edge, increasing carrier lifetimes and mobilities
by an order of magnitude.
We also predict energy distributions of carriers excited upon optical absorption,
and find that they are dominated by direct transitions rather than indirect
phonon-assisted transitions for all relevant photon energies.

\begin{figure} 
\includegraphics[width=\columnwidth]{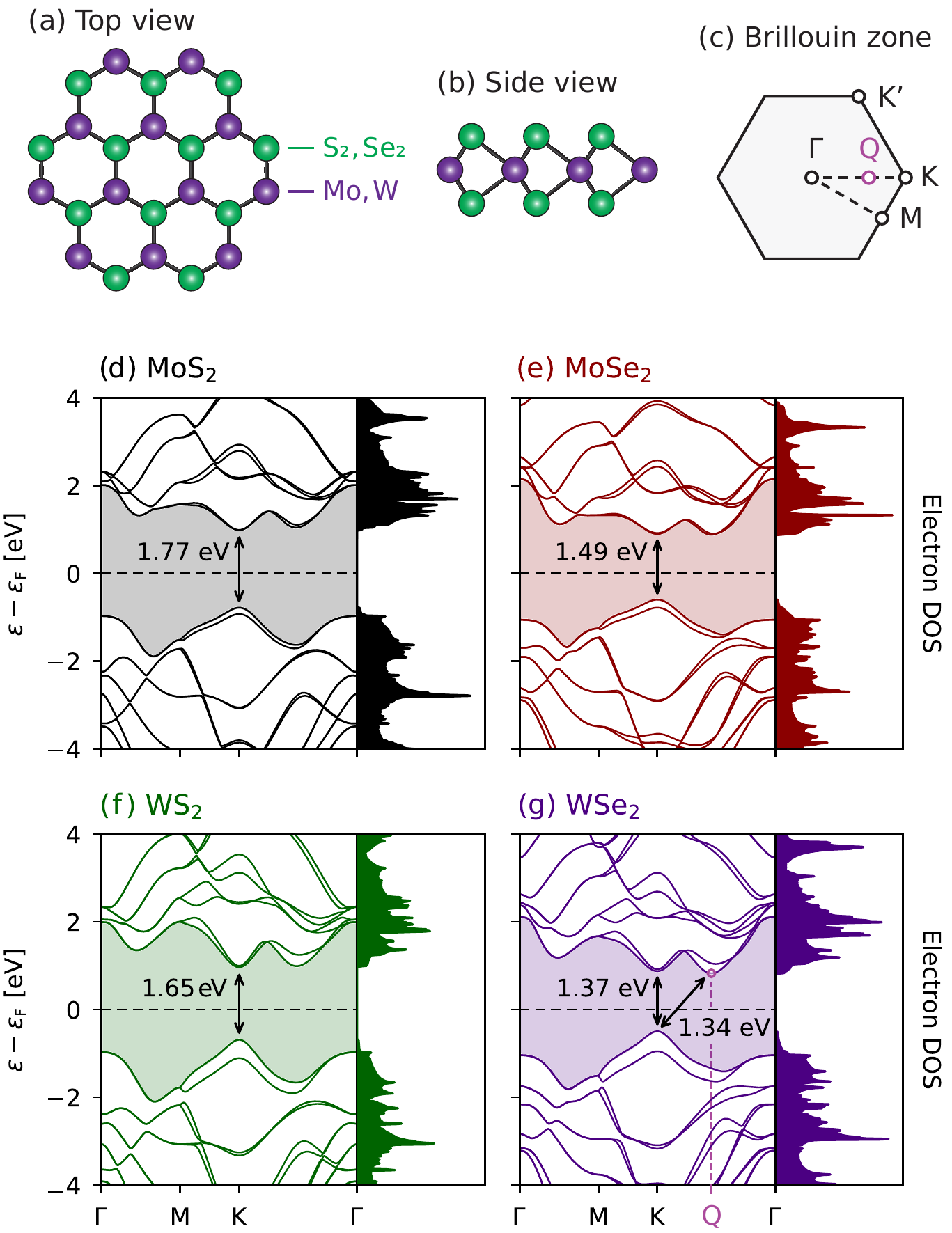}
\caption{Monolayer TMDC structure and electronic band structure.
(a,b) TMDC crystal structure and  (c) corresponding Brillouin zone. The $K$ and
$K'$ points are energetically degenerate but inequivalent due to a lack of an
inversion center. (d--g) Calculated electronic band structure and density of states
for each monolayer, with band gaps annotated. Three of the four monolayers are
predicted to be direct band gap semiconductors, except for WSe\sub{2} which
is predicted to have an indirect band gap with the conduction edge
at the $Q$ point, consistent with experiment~\cite{WSe1,WSe2}.
\label{fig:bandstructure}}
\end{figure}

We start with first-principles electronic structure calculations
of the four TMDC monolayers considered here, sulfides and selenides
of tungsten and molybdenum, all of which adopt the hexagonal 
crystal structure illustrated in Fig.~\ref{fig:bandstructure}.
Corresponding electronic band structures and densities of states
are shown in Fig.~\ref{fig:bandstructure}d--g. 
We find that WSe\sub{2} is predicted to have an indirect gap,
with its conduction band edge at the $Q$ point rather than the $K$ point
(see Fig.~\ref{fig:bandstructure}c), consistent with experimental findings~\cite{WSe1, WSe2},
while the remaining three materials have a direct gap at the $K$ point.
All materials are mechanically stable as indicated by no imaginary
frequencies in the calculated phonon band structures (see SI).

\begin{figure} 
\includegraphics[width=\columnwidth]{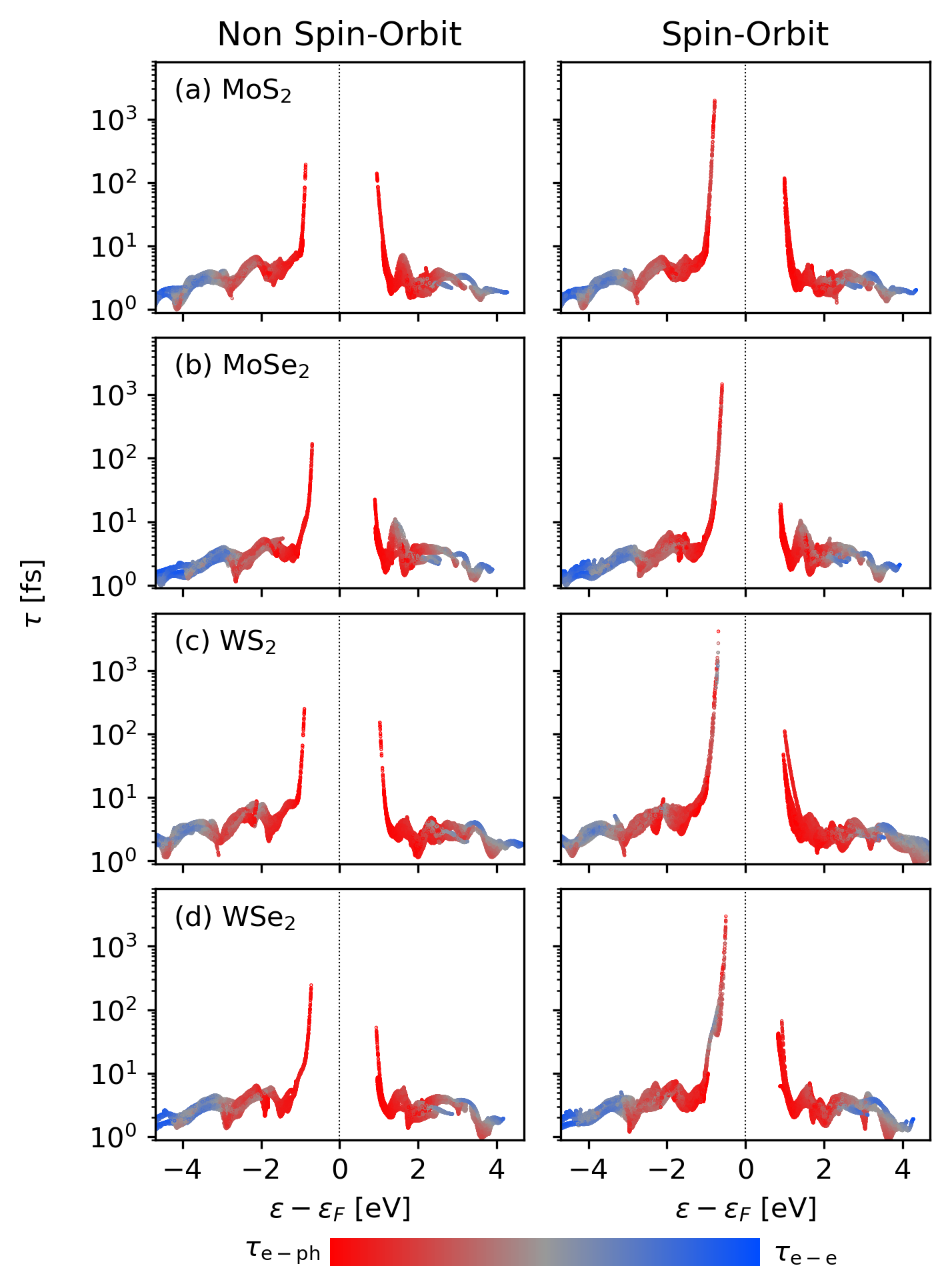}
\caption{Total scattering lifetimes for hot carriers near the Fermi
level ($T = \text{298~K}$). Calculations are performed with (right) and without
(left) spin-orbit coupling.  The color bar indicates the relative contributions
to scattering from from electron-phonon (red) and electron-electron (blue)
interactions. Lifetimes are enhanced by over an order of magnitude at the
valence band edge when spin-orbit effects are considered for each monolayer, due
to spin-valley locking.
\label{fig:lifetimes}}
\end{figure}

Next, we use first-principles calculations of electron-electron and electron-phonon
scattering rates to predict the net carrier lifetimes, shown as a function
of carrier energy in Fig.~\ref{fig:lifetimes}.
The relative contributions of the two scattering mechanisms to the total
scattering rate, $\tau_{\vec{k}n}^{-1} = (\tau\super{e-e}_{\vec{k}n})^{-1}
+ (\tau\super{e-ph}_{\vec{k}n})^{-1}$, are shown using the color scale.
Electron-phonon scattering (red) dominates the net scattering near
the band edges, while electron-electron scattering picks up further
from the band edges due to a quadratically increasing phase space for scattering.
The scattering times near the band edges are the longest,
because the phase space for electron-phonon scattering is proportional
to the density of states near the carrier energy, which vanishes at the band edges.

{\bf{Spin-Valley Locking Captured from First Principles.}}
Overall, in Fig.~\ref{fig:lifetimes}, the predicted scattering times with and without
spin-orbit coupling are qualitatively similar throughout, and quantitatively
similar far from the band edges.
However, especially near the valence band edge, spin-orbit coupling dramatically
alters the electron-phonon scattering rate, increasing the net hole
lifetime by about an order of magnitude in all four monolayers.
At room temperature (298~K), maximum hole lifetimes in MoS\sub{2} and MoSe{2}
are $\sim$2~ps, while they exceed 4~ps in the tungsten TMDCs.

\begin{figure} 
\includegraphics[width=0.75\columnwidth]{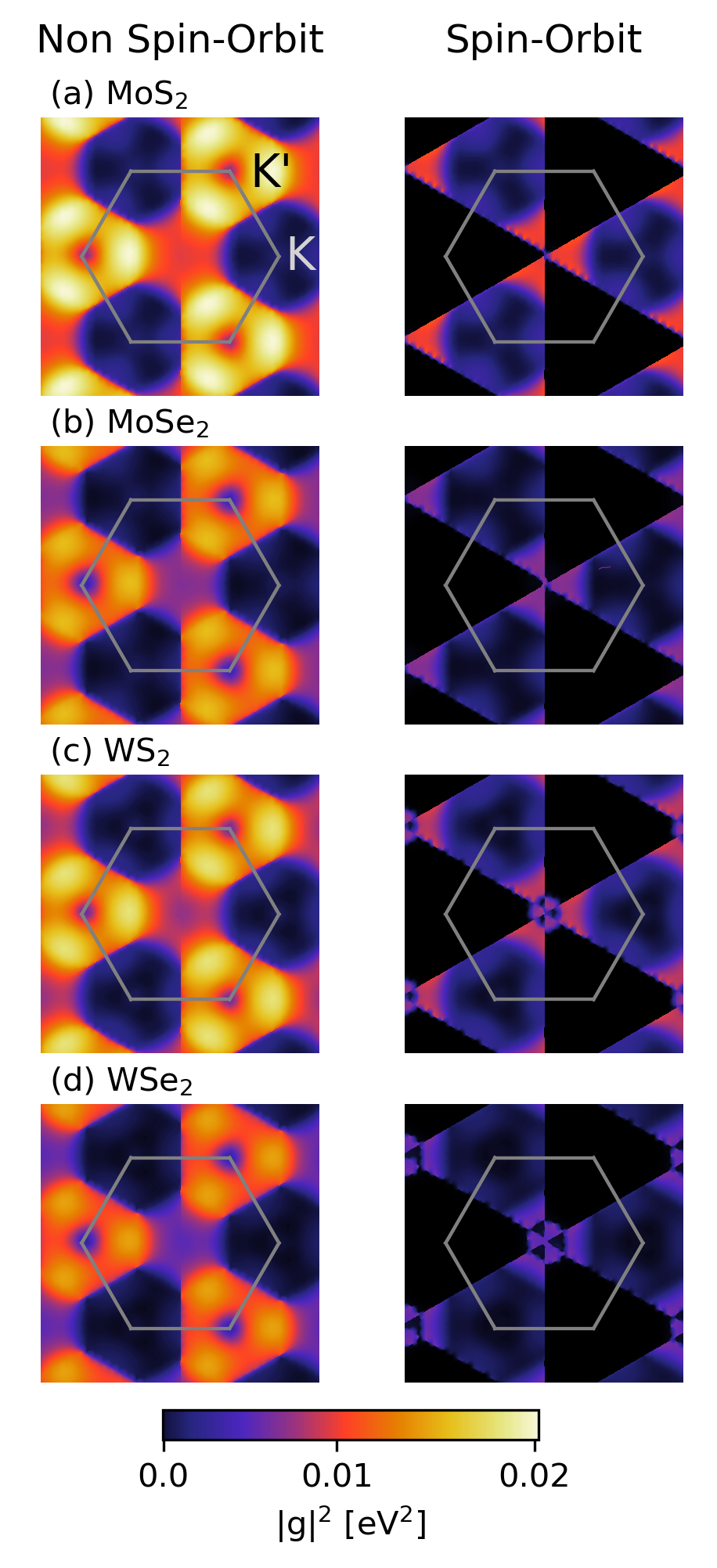} 
\caption{Electron-phonon coupling for valence-band-edge carriers.
Electron-phonon matrix element squared and summed over phonon modes,
between the valence-band edge state at the $K$ point and the highest 
valence band state for each wave-vector $\vec{k}$ in the Brillouin zone (BZ),
both with and without spin-orbit coupling for each of the four monolayer TMDCs.
Without spin-orbit coupling, the matrix element is mostly non-zero throughout the BZ,
while with spin-orbit coupling, half the BZ goes dark with essentially zero
electron-phonon coupling to the highest occupied band.
This is because these transitions would involve 
final electronic states with opposite spin.
The strong spin-orbit splitting causes
these states to be inaccessible via electron-phonon interactions.
This results in spin-valley locking and much higher carrier lifetimes. 
\label{fig:ePhcoupling}}
\end{figure}

To understand the reason for lifetime enhancement due to spin-orbit coupling,
Fig.~\ref{fig:ePhcoupling} shows the calculated electron-phonon matrix elements
(squared and summed over all phonon modes) connecting a state at 
the valence band edge at the $K$ point of the Brillouin zone (BZ)
with the highest valence band states at all other points of the BZ.
Note that without spin-orbit coupling, the results are six-fold symmetric
and there is strong electron-phonon coupling between the $K$ and $K'$ points,
which are inequivalent due to the absence of inversion symmetry.
In this case, there are two degenerate spin states at each of the
$K$ and $K'$ valleys, and phonons strongly couple the states with the same spin.
Since these are at the same energy, these states are accessible
for electron-phonon scattering and result in intervalley scattering
that limits the lifetime(s) of the carriers.

Spin-orbit coupling completely changes this picture, as shown in Fig.~\ref{fig:ePhcoupling}.
The two spin states in each of the $K$ and $K'$ valleys are no longer degenerate,
and the energy split occurs in the \emph{opposite} direction at the two valleys.
Consequently the valence band edge at the $K$ and $K'$ points have opposite spins,
and the intervalley scattering between these equal energy states must involve
a spin flip~\cite{Xu:2014}, which has an extremely small matrix element.
This manifests in Fig.~\ref{fig:ePhcoupling} as an entire half of the BZ
centered on the intervalley scattering process has an essentially
zero electron-phonon matrix element.
The phonon states which coupled the spin-degenerate electronic states in the 
case without spin-orbit coupling have now been split away to an energy inaccessible
at room temperature for electron-phonon scattering.
This forces the carriers of a given spin to remain locked to a given valley, and
this spin-valley locking produces the sharp increase in the electron-phonon lifetime
of holes near the valence band edge for all four TMDCs in Fig.~\ref{fig:lifetimes}.
On the other hand, the spin-orbit split at the conduction band edge is negligible
and there is hence no electron lifetime enhancement compared to the non spin-orbit case.

As shown above, the rate of intervalley scattering determines
the valence-band-edge carrier lifetimes. These carrier lifetimes
are therefore the time for which holes remain locked to a valley,
effectively the retention time of valley information in valleytronic devices.
Despite a number of experimental investigations using a variety of methods,
the valley retention times have not yet been conclusively determined. 
Using exactly the same first-principles methodology as above,
we also calculate the band-edge hole lifetimes at $T = 50$~K,
where the lifetimes are expected to be longer (and more useful
for valleytronics) due to lowered  phase space for scattering.
We find the valley retention time to be 57~ps and 67~ps for
MoS\sub{2} and MoSe\sub{2}, and 62~ps and 138~ps for 
WS\sub{2} and WSe\sub{2}. The relative lifetime
values among the four monolayers
correlate with the strength of the electron-phonon coupling
seen in Fig.~\ref{fig:ePhcoupling}. This lifetime trend
also correlates with the predicted energy difference between
the $\Gamma$ and $K$ valleys at the valence edge.

The methods used to calculate these lifetimes only capture processes in
an atomically perfect crystal, ignoring potential
interactions with defects and substrates,
and therefore represent a best-case scenario for monolayer TMDC valley lifetimes.
Unambiguous experimental determination of the limiting valley retention times
is challenging precisely because it is impossible to disentangle substrate 
interactions and dopant/defect effects.
Additionally, excitonic and other multi-particle effects complicate
signatures from optical measurements, all of which we exclude
in our theoretical predictions above.
Consequently, we predict the best-case valley lifetime in monolayer TMDCs
at 50~K to exceed the 100 ps scale for WSe\sub{2}, and to be roughly
on the same order of magnitude for the others. 

\begin{figure} 
\includegraphics[width=0.9\columnwidth]{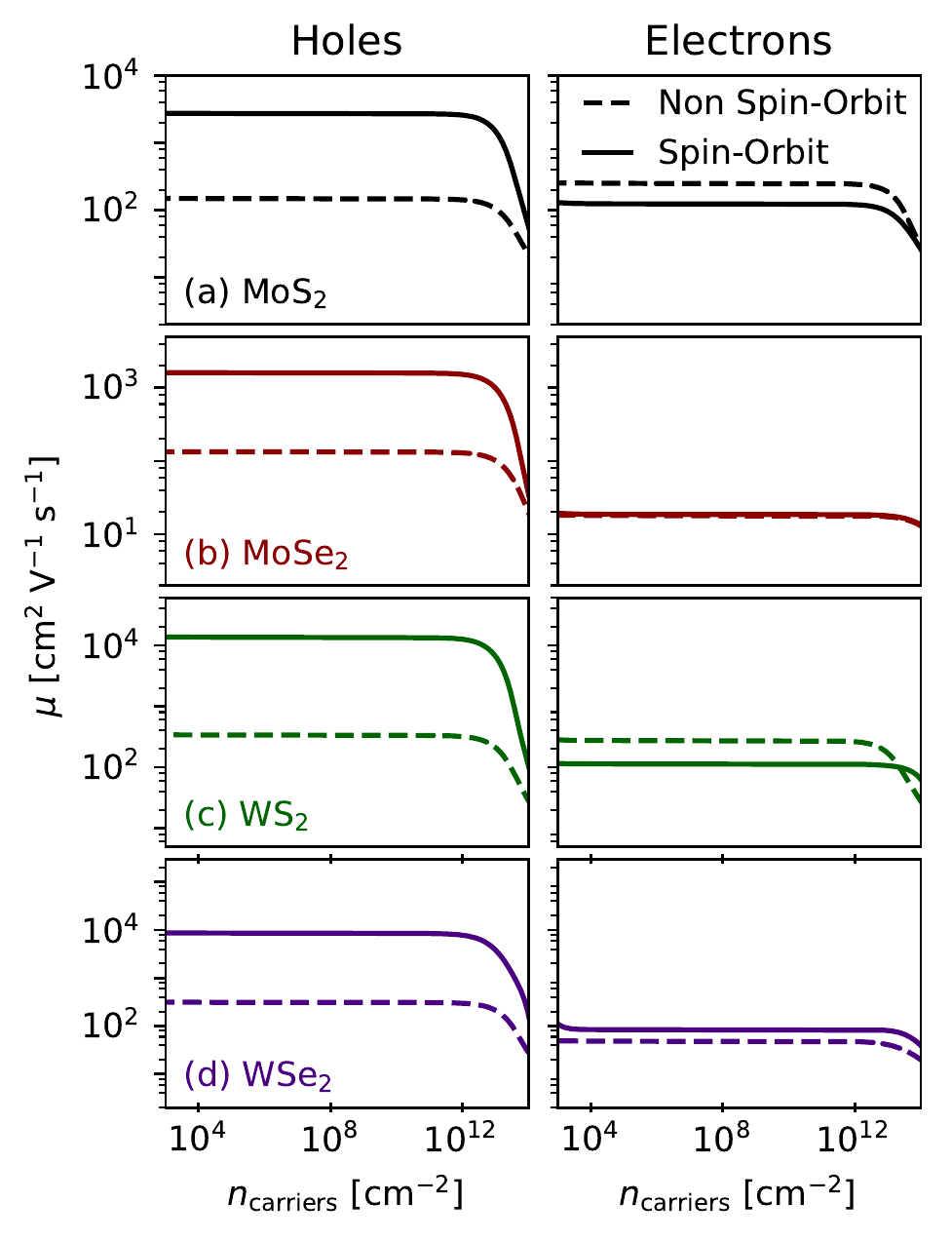} 
\caption{Hole and electron mobility versus carrier density at $T = 298$~K.
Hole mobility is enhanced by spin-valley locking by over an order of magnitude
across all four TMDCs in the calculations including spin-orbit coupling,
compared to those that do not include it.
The corresponding effect on electron mobility is less pronounced,
where instead lowering of the $Q$ valley in the spin-orbit calculations
introduces low-energy scattering and decreases the mobility.
WSe\sub{2} is an exception where the $Q$ point is sufficiently
isolated from $K$ at the conduction edge, such that this scattering is suppressed
and the electron mobility increases.
\label{fig:mobility}}
\end{figure}

{\bf{Spin-orbit Enhanced Hole Mobility.}}
The enhancement of band-edge carrier lifetimes due to suppression
of intervalley scattering should result in a corresponding increase
in the carrier mobility due to spin-orbit coupling
(since mobility $\mu = e\tau/m^\ast \propto \tau$ in Drude theory).
Fig.~\ref{fig:mobility} shows the intrinsic carrier mobility
due to electron-phonon scattering as a function of carrier density,
both for electrons and holes, with and without spin-orbit coupling.
As expected, the hole mobility is enhanced by over an order of magnitude 
when spin-orbit coupling is considered, exactly as was the case for carrier lifetimes.
Previous theoretical predictions of hole mobilities in these materials,\cite{Kaasbjerg:2012,Li:2013, Ma:2014, Jin:2014}
are all below 1000 \mobUnit because they do not include spin-orbit interactions;
our non-relativistic calculations are in good agreement with these previous studies.
The trend in hole mobility is consistent with the corresponding lifetimes,
with the largest hole mobility in WS\sub{2} $\sim 10^4$~\mobUnit.

On the other hand, electron mobility is less drastically
affected by spin-orbit coupling. The differences
can be explained based on the band structures calculated
with and without spin-orbit coupling.
In the spin-orbit case, the conduction band edge is found
to be nearly degenerate between the $K$ and $Q$ points 
(see Fig.~\ref{fig:bandstructure}). This introduces
low-energy scattering between these two valleys. Meanwhile,
non-relativistic calculations find the $K$ and $Q$ valleys are 
more energetically separated, making intervalley scattering
via phonons less accessible. Consequently, electron
mobility is predicted to decrease due to spin-orbit coupling
for some TMDCs, as shown in Fig.~\ref{fig:mobility}. 
However, in the case of WSe\sub{2}, the energy difference 
between the $K$ and $Q$ conduction valleys is larger, and
spin-orbit coupling increases the electron mobilities instead.

Experimentally-measured mobilities in monolayer TMDCs
are smaller than theoretical predictions
due to several inherent non-idealities including
substrate effects, trapped impurities, air-borne adsorbates,
and overall sample quality including defects and grain size~\cite{Yu:2017,Yi:2017}. 
Our predictions shed light on the underlying physics of carrier lifetimes
and transport in 2D TMDCs, underscoring the importance of spin-valley locking
not just in valleytronics, but also in overall charge transport for
electronic applications such as in field-effect transistors.

\begin{figure} 
\includegraphics[width=\columnwidth]{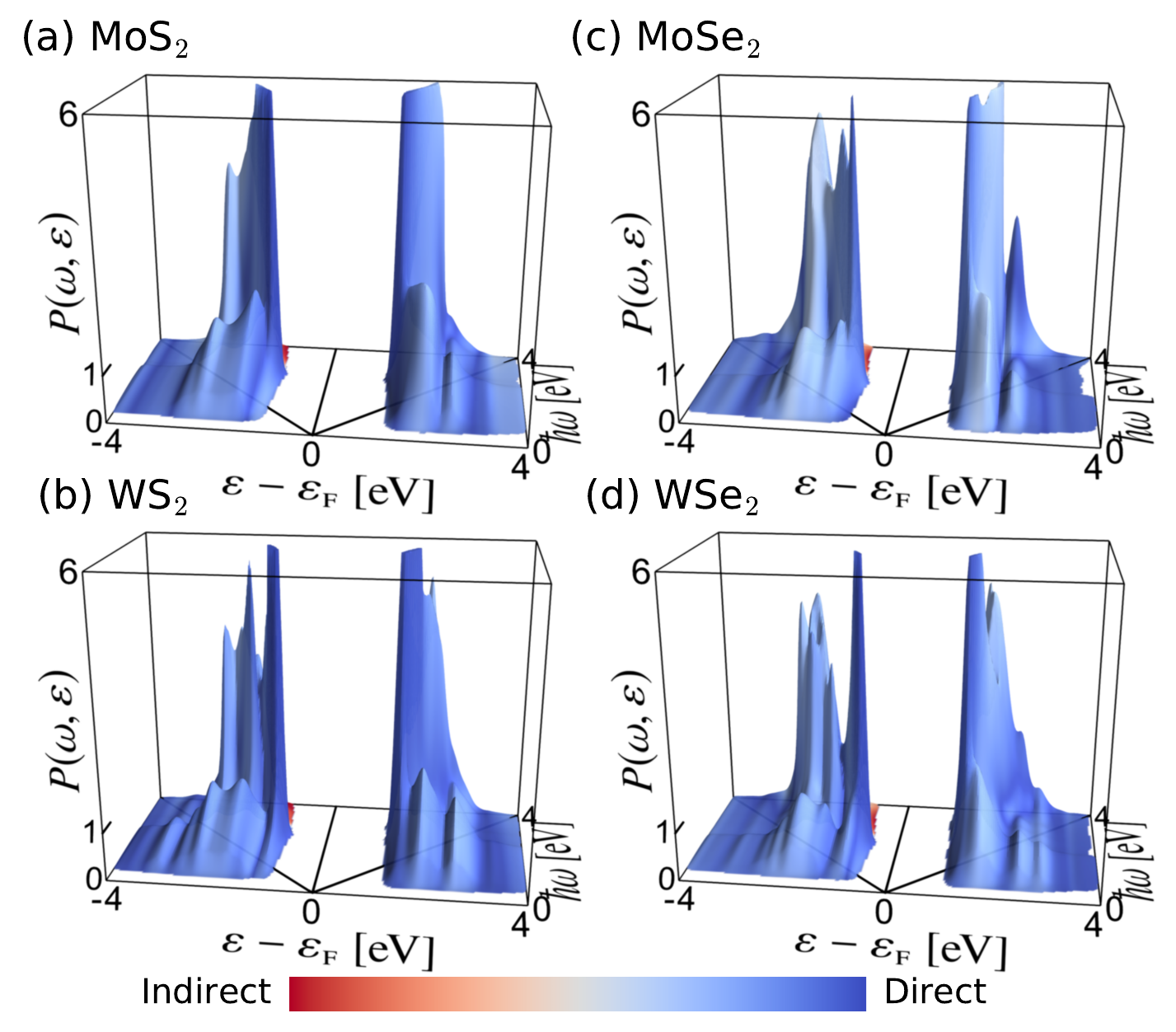} 
\caption{Hot carrier energy distributions as a function of optical excitation energy.
The distribution at each photon energy is normalized such that a flat distribution would yield 1,
while the color scale indicates the contribution from direct and phonon-assisted processes.
Direct transitions dominate hot carrier generation for each of these monolayers,
even in the case of the (slightly) indirect-gapped WSe\sub{2}.
\label{fig:carriers}}
\end{figure}

Finally, we investigate the energy distributions of carriers that are excited
upon optical absorption in these materials, accounting for both direct
and phonon-assisted transitions using our previously established
first-principles methodology.\cite{Brown:2016, AdvOptMat, Coulter:2018, 
NitrideCarriers, Papadakis:2018fk, NatCom, PhysRevB.94.075120, PhysRevB.97.195435}
Fig.~\ref{fig:carriers} show that direct transitions dominate carrier generation
in all these materials, as expected for direct gap semiconductors where the
band gap and optical gap are equal so that direct transitions are always allowed.
However, this is also the case for the indirect-gap WSe\sub{2} because
the small energy difference of the conduction band edges at the $K$ and $Q$ points
results in a small difference between the band gap and optical gap, making it
behave essentially as a direct band gap semiconductor.

{\bf{Conclusions.}} 
We use first-principles calculations of carrier dynamics in monolayer TMDCs
with an \emph{ab initio} treatment of electron-electron and electron-phonon interactions
to elucidate the critical effect of spin-orbit coupling in these materials.
Our results highlight the importance of spin-valley locking of holes near the
valence band edge at the $K$ and $K'$, a consequence of spin-orbit coupling,
on carrier lifetimes, valley retention time and charge transport.
In particular, we find that spin-orbit coupling enhances the hole lifetimes
and mobilities by an order of magnitude in all four materials considered here.
Electron lifetimes and mobilities are less affected in comparison,
due to much smaller spin-orbit coupling effects near the conduction band edge.
We predict the ideal valley relaxation time in these materials at
a lower temperature of 50~K to be at the 100~ps scale,
with the largest value for WSe\sub{2} $\sim$140~ps exceeding
the others by about a factor of two.
While our results focus on monolayer systems, multilayer and heterostructured TMDC systems
with strong spin-orbit coupling should also have similar spin-valley locking physics,
which necessitate a careful analysis of the intervalley scattering mechanisms, fully
accounting for the effect of phonons from a self-consistent spin-orbit coupling perspective.

\section*{Associated Content}
{\bf{Supporting Information}.} Phonon dispersion and additional details
regarding calculations of optical response.

\section*{Author Information}
The authors declare no competing financial interests.

\section*{Acknowledgments}

This research used resources of the National Energy Research Scientific
Computing Center, a DOE Office of Science User Facility supported by the Office
of Science of the U.S. Department of Energy under Contract No. DE-AC02-05CH11231,
as well as resources at the Research Computing Group at Harvard University.
TC acknowledges support from the Danish
Council for Independent Research (Grant No. DFF-6108-00667).
RS acknowledges start-up funding from the Materials Science and Engineering
department at Rensselaer Polytechnic Institute.
PN acknowledges start-up funding from the Harvard John A.
Paulson School of Engineering and Applied Sciences.

\section*{Methods}
\small
{\bf{Computational Details.}}
We used the open-source JDFTx density-functional theory software
for structural relaxation, electronic band structure, phonon
and electron-phonon matrix element calculations.\cite{JDFTx}
We carried out all calculations each with relativistic
and non-relativistic ultrasoft pseudopotentials~\cite{USPP}
to investigate the effect of spin-orbit coupling.
For the exchange-correlation functional, we used the PBEsol
generalized-gradient approximation~\cite{PBEsol}, which yielded
relaxed lattice constants within 1\% of experimental values.
To eliminate effect of periodic images in the out-of-plane direction,
we used truncated Coulomb interactions throughout for these 2D materials.\cite{TruncatedEXX}

All electronic calculations employed a $18\times 18\times 1$ 
$\Gamma$-centered $\vec{k}$-point mesh for BZ sampling
with a plane-wave energy cutoff of 30 Hartrees.
Phonon properties were calculated from symmetry-irreducible
perturbations in a $6\times 6\times 1$ supercell, also done
both with and without self-consistent spin-orbit coupling.
Directly calculating electron-phonon scattering properties in DFT
is expensive due to the energy mismatch between electron and phonon
scales necessitating extremely fine BZ sampling.
Consequently, we convert all electron, phonon and electron-phonon
properties calculated at the above `coarse' BZ meshes
to a basis of maximally-localized Wannier functions~\cite{MLWFmetal}
(starting from transition metal $d$ and chalcogen $p$ trial orbitals).
We then interpolate these properties~\cite{Giustino:2007,Brown:2016}
to substantially finer electron $k$ and phonon $q$ meshes
with $\sim 1000$ points per dimension ($\sim 10^6$ total),
used for all carrier scattering and optical response properties described below.

{\bf{Carrier Lifetimes.}}
The total carrier lifetime is determined by electron-electron 
and electron-phonon scattering, with the total scattering rate
given by Matthiessen's rule:
\begin{equation}
\tau^{-1}_{\vec{k}{n}}
	= \Big( \tau\super{e-e}_{\vec{k}{n}} \Big)^{-1}
	+ \Big( \tau\super{e-ph}_{\vec{k}{n}} \Big)^{-1}
\end{equation}
for electrons of each band $n$ at wave-vector $\vec{k}$ in the 
two dimensional BZ. The electron-electron scattering rate is calculated
from the imaginary part of the carrier self-energy, given 
by\cite{Brown:2016,eeLinewidth}:
\begin{multline} 
\Big(\tau\super{e-e}_{\vec{k}{n}}\Big)^{-1}
= \frac{2\pi}{\hbar} \int\sub{BZ} \frac{\mathrm{d}\vec{k}'}{(2\pi)^2}
	\sum_{n'} \sum_{\vec{G}\vec{G}'}
	\tilde{\rho}_{\vec{k}'n',\vec{k}n}(\vec{G})
	\tilde{\rho}_{\vec{k}'n',\vec{k}n}^\ast(\vec{G}')
\\ \times
	\frac{1}{\pi}\Im W_{\vec{G}\vec{G}^\prime}
	(\vec{k}^\prime-\vec{k}, \varepsilon_{\vec{k}n}
		-\varepsilon_{\vec{k}'n'}).
\label{eqn:tauInv_ee} 
\end{multline}
The above expression is essentially the imaginary part
of the interaction of the one-particle electronic density
matrices $\tilde{\rho}_{\vec{k}'n',\vec{k}n}(\vec{G})$
through the dynamically screened Coulomb interaction,
$W_{\vec{G}\vec{G}^\prime}(\vec{k}'-\vec{k}, \omega)$
evaluated within the random phase approximation.
This calculation is performed directly in the plane-wave
basis of reciprocal lattice vectors $\vec{G}$ and $\vec{G}'$,
and involves a sum over a full second set of electronic states ($\vec{k}'n'$).
See Ref.~\citenum{Brown:2016} for further details.

We calculate the electron-phonon scattering rate using
Fermi's Golden rule~\cite{Brown:2016}:
\begin{multline} 
\left(\tau\super{e-ph}_{\vec{k}{n}}\right)^{-1}
= \frac{2\pi}{\hbar} \sum_{n'\alpha\pm} \int\sub{BZ}
	\frac{\Omega\mathrm{d}\vec{k}'}{(2\pi)^2}
	\delta(\varepsilon_{\vec{k}'n'} - \varepsilon_{\vec{k}n}
		\mp \hbar\omega_{\vec{q}\alpha})
\\\times
	\left[ n_{\vec{q}\alpha} + \frac{1}{2} \mp \left(\frac{1}{2}
		- f_{\vec{k}'n'}\right)\right]
	\left| g^{\vec{q}\alpha}_{\vec{k}'n',\vec{k}n} \right|^2,
\label{eq:tauInv_ePh} 
\end{multline} 
where $\varepsilon_{\vec{k}n}$ and $f_{\vec{k}n}$ are energies
and Fermi occupations of electrons at wave-vector $\vec{k}$
in band $n$, $\omega_{\vec{q}\alpha}$ and $n_{\vec{q}\alpha}$
are angular frequencies and Bose occupations of phonons
at wave-vector $\vec{q}$ with polarization index $\alpha$,
and $g^{\vec{q}\alpha}_{\vec{k}'n',\vec{k}n}$ is the electron-phonon
matrix element coupling them to final electronic state ($\vec{k}'n'$)
(a three-vertex in the diagrammatic picture)
with $\vec{q} = \vec{k}'-\vec{k}$ by momentum conservation.
Above, the summation over $\pm$ accounts for phonon 
emission and absorption processes.

{\bf{Mobility.}}
For calculating carrier mobilities, we first evaluate carrier
momentum relaxation times due to electron-phonon scattering,
\begin{multline} 
\left(\tau^p_{\vec{k}{n}}\right)^{-1}
= \frac{2\pi}{\hbar} \sum_{n'\alpha\pm} \int\sub{BZ}
	\frac{\Omega\mathrm{d}\vec{k}'}{(2\pi)^2}
	\delta(\varepsilon_{\vec{k}'n'} - \varepsilon_{\vec{k}n}
		\mp \hbar\omega_{\vec{q}\alpha})
\\\times
	\left[ n_{\vec{q}\alpha} + \frac{1}{2} \mp \left(\frac{1}{2}
		- f_{\vec{k}'n'}\right)\right]
	\left| g^{\vec{q}\alpha}_{\vec{k}'n',\vec{k}n} \right|^2
\\\times
	\left( 1 - \frac{\vec{v}_{\vec{k}n}\cdot\vec{v}_{\vec{k}'n'}}
		{|\vec{v}_{\vec{k}n}| |\vec{v}_{\vec{k}'n'}|} \right),
\label{eq:tauInv_ePhP} 
\end{multline} 
which is identical to (\ref{eq:tauInv_ePh}) except for an
additional final factor accounting for the scattering angle
between initial and final electron band velocities $\vec{v}_{\vec{k}n}$
(defined by $\vec{v} \equiv \partial\varepsilon/\partial\vec{k}$).
Then, we calculate the mobility by solving the linearized
Boltzmann equation using a full-band relaxation-time
approximation,\cite{Brown:2016,NitrideCarriers,Gunst:2016}
\begin{equation} 
\bar{\mu}(\varepsilon_\textsc{f}) = \frac{e}{|n(\varepsilon_\textsc{f})|}
	\sum_{n}\int\sub{BZ}\frac{g_s\mathrm{d}{\vec{k}}}{(2 \pi)^2}
	\frac{\partial f_{\vec{k}n}(\varepsilon_\textsc{f})}{\partial \varepsilon_{\vec{k}n}}
	( \mathbf{v}_{\vec{k} n} \otimes \mathbf{v}_{\vec{k} n} )
	\tau^p_{\vec{k} n},
\label{eq:mu}
\end{equation}
where the Fermi function derivative selects out carriers that contribute
to transport at a particular doping level specified by Fermi level
position $\varepsilon_\textsc{f}$, and where $g_s$ ($= 1$ with and $= 2$
without spin-orbit coupling) is the spin-degeneracy factor.
Above, the Fermi-level dependent carrier density is defined as
\begin{equation*}
n(\varepsilon_\textsc{f}) = \sum_{n}\int\sub{BZ}\frac{g_s\mathrm{d}{\vec{k}}}{(2 \pi)^2}
	f_{\vec{k}n}(\varepsilon_\textsc{f}) - n_0,
\end{equation*}
where $n_0$ is the number density of carriers in the neutral DFT calculation;
this is positive for $n$-type semiconductors with a net electron density
and negative for $p$-type semiconductors with a net hole density.
By varying the Fermi level position from near the valence band edge
to near the conduction band edge, we trace out the hole mobility
(when $n<0$) and then the electron mobility (when $n>0$) as a
function of carrier density $|n|$ as shown in Fig.~\ref{fig:mobility}.

\makeatletter{}

\end{document}